# Identical Particles in Quantum Mechanics

Andrea Lubberdink[1]
Institute for History and Foundations of Science,
Dept. of Physics and Astronomy, Utrecht University

## 1. The Problem of Indistinguishable Identical Particles in Quantum Mechanics (h-particles)

Consider the state of all protons in the universe. If we define the state of one proton by the partial trace of the total symmetrized state to one of the component spaces, $H_i$ , then all proton states are identical. So, defined in this way, all protons in the universe are in exactly the same state, all protons are indistinguishable.

But that does not correspond to the way the term proton is actually used by physicists. For example, they talk about: "colliding protons at LHC in Cern".
This distinguishes protons at LHC in Cern from all other protons in the universe.
Or:
"The ALSEP determined that more than 95% of the particles in the solar wind are electrons and protons…"
This distinguishes protons in the solar wind from all other protons in the universe.

We can find countless similar examples of the use of the term proton in which certain protons are being distinguished from the many other protons in the universe. When it comes to experiments and observations, protons are always somehow localized, e.g. in Cern, in the solar wind or in an H atom. In observations or experiments, the considered protons are always in some way distinguished and pointed at, individually or as part of a selection of protons.

Because of the above, which of course holds not only for protons but for all identical particles, I think it is wrong to define a particle as being in a state which is the partial trace of the total state to one of the component spaces of the total tensor product Hilbert space. It seems obvious to me, that if we use the term 'particle' in QM, it should be defined in such a way that in the classical limit the quantum particle state becomes the state of a classical particle.

The scientific use of the term particle originates in classical mechanics, where the very idea of particles is that they can be pointed at! A classical particle is always individualized by having a unique position at each moment. Defining particles in quantum mechanics, in such a way that they cannot be pointed at, not even in the classical limit, seems a strange thing to do. Why call something a 'particle', when it has no connection to the particles we know from classical mechanics? Of course we can give a name to a theoretical term in QM, but if we call it 'particle' we should be very careful, because this suggests a connection to the particles we already have an intuition of, resulting from daily life and classical mechanics.

Using the term 'particle' in such a way that they are all in the same state gives rise to philosophical worries (see e.g., [1], … ,[5]). The indistinguishability may be seen as one of the mysteries or peculiarities of QM. However, I argue that it is merely the result of a wrong definition of the term

[1] Email address: andrea@andrealubberdink.nl

‘particle’. In the following I will call these ‘particles’ that are characterized by partial trace states: h-particles.

## 2. Indistinguishable Identical Particles in Classical Mechanics, analogous to those in QM

Below, I will indicate how we can define classical indistinguishable identical particles, which are analogous to the indistinguishable h-particles in quantum mechanics. This will show that indistinguishability of identical particles is not a typical quantum mechanical phenomenon. And also it will show that indistinguishability of identical particles is not an inevitable result of the symmetry postulate. It is the result of a wrong definition of the term “particle”.

We don’t ordinarily use any symmetry postulate in classical mechanics, but we may introduce such a postulate without any change in empirical content. The resulting theory, which I will outline below, will be denoted here as ‘the modified classical mechanics’.

In the usual classical mechanics the state of an n-particle system is one point in x-p phase space. In the modified classical mechanics with the symmetry postulate the state of an n-particle system is the collection of n! points in phase space, where the n! points correspond to all permutations of the particles.

> To see how this works, just consider the simple example of only two identical particles and draw only the x-axis for both particles, that is the $x_1$ and $x_2$ axis. If the two particles are at positions 2 and 3 then draw the point: $(x_1, x_2) = (2,3)$. This (together with the p-values, $p_a$ and $p_b$, which we will leave out here for simplicity) would give the state of the two particles in usual classical mechanics. Now we add the point: $(x_1, x_2) = (3,2)$, which could also give the state of the two particles in the usual classical mechanics. In the usual classical mechanics we would have to choose one of these points to represent the state of the system. In our modified classical mechanics we now define the collection of both points as the state of the system, i.e. the state of system is: $(x_1, x_2) = \{(2,3), (3,2)\}$. This state is symmetrical, i.e. if the labels of the x-axes are interchanged, the state remains the same.

We can now define the state of one particle, particle i, by ‘projecting’ the collection of points in phase space to the $x_i$, $p_i$-axes. The result is a collection of the n! x, p -values. These are the same for all i.

> In our example it would be that both particles are in the same state being: {2,3}, or, without forgetting p, the state would be: $\{(2,p_a),(3,p_b)\}$

So, in our version of classical mechanics all particles are in the same state, namely the projection of our generalized n!-points state onto one of the individual axes. This is completely analogous to what we have in QM, where all h-particles are in the same state, namely the partial trace to $\mathcal{H}_i$.

> Just consider a two particle system in QM in the symmetrical state: $1/\sqrt{2}\ \{|\phi_1, \phi_2\rangle + |\phi_2, \phi_1\rangle\}$ .
> This corresponds to the state {(2,3), (3,2)} of our example above.
>
> Taking the partial trace in QM gives: $W_1=W_2= ½\ \{|\phi_1\rangle\langle \phi_1| + |\phi_2\rangle\langle \phi_2|\}$
> The partial tracing corresponds to the `projecting’ in the above example.

The analogy with quantum mechanics is clear.

## 3 Identical Particles in Quantum Mechanics (q-particles)

We have seen that in classical mechanics we can also introduce a symmetry postulate. We then get a modified classical mechanics, with exactly the same empirical results as usual classical mechanics. Indeed, although the symmetry postulate is not necessary in classical mechanics, it is appropriate and elegant to have it in, because of symmetry considerations.

We have seen that if we define classical particle states by means of projecting from the symmetrized state of the total system, the result is that all 1-particle states are the same, so that the particles have become indistinguishable. So we see, that defining classical particle states in this way does not result in the different particle states we started out with. Apparently we must not associate the index of the x and p axis in phase space with the label of a particle. We get the right particle states if we just associate a particle to each different position value.

In analogy, if we define particle states in QM by means of partial tracing from the (anti) symmetrized state of the total system, the result is that all 1-particle states are the same, so that all particles (h-particles) are indistinguishable. The analogy now leads us to propose that we must not define particle states in QM by means of partial tracing, i.e., we must not interpret the index, labeling the Hilbert spaces, as a particle index. What we must do is label each hump in configuration space. In other words, we have to separate different position distribution functions and call these particle states. Particles defined this way, I denote as q-particles.

This seems to be exactly what we have learned not to do, what we have learned to be a big mistake, a 'beginners mistake'. Because we know that the position distribution functions cannot be separated uniquely if they overlap, so q-particles will then not be uniquely defined. So be it. They can be uniquely defined if the functions don't overlap[2], as will be in the classical case. And also, approaching the classical limit, the definition of the particles will make more and more sense. [3]

Looking back, we may note that the way the term particle is actually mostly used by physicists, refers to q-particles. In literature, the term particle sometimes refers to q-particles and sometimes to h-particles. For example, in statistical mechanics, the particle states refer to q-particles. Indeed, we couldn't do statistics on particle-states if they were all equal.
q-particles correspond to the intuitive particle concept. We need this concept, because this is how we perceive the world. We live and learn by means of distinguishing things (which consist of particles). We point at them. Of course, in general, particles in QM cannot be sharply pointed at, because they don't have an exact position, but to make the connection between theory and observations, we need a particle concept, which at least approximates the classical particle concept.

The inadequacy of h-particles can be clearly seen from the fact that they don't ever approximate classical particles. They remain indistinguishable, even if I can almost 'see' the particles. Even if I cannot see the individual particles I can see my table, for example. I can point at it. How can it consist only of particles whose position is 'spread out' all over the universe? Please note that I don't object to the concept of indistinguishabily. I do object to indistinguishability of all identical particles in the universe. This is because all things, (even we, ourselves) consist of nothing but identical particles. We wouldn't be able to distinguish anything, (not even ourselves from one another) if it was built only of indistinguishable particles.

---

[2] I have given a mathematical proof of this in my Master Thesis, written in 1998 (in Dutch). The Thesis can be found at: www.andrealubberdink.nl/identicalparticles/English.html . Also, the statements and arguments given in this paper are discussed more extensive in the Thesis.

[3] The reasoning I sketched above for defining identical particle states, can be broadened to defining the states of subsystems, consisting of n identical particles. We want to be able to point to a subsystem, especially since all n-particle systems actually are subsystems of the ultimate N-particle system, which consists of all particles in the universe. Even if the q-particles in the subsystem may be so much entangled that it would make no sense to define their states, it is important to define the state of the subsystem, as being different from other subsystems. This, again, will not be achieved by partial tracing.

In actual practice, this problem with the indistinguishabily of h-particles or systems of h-particles is usually by-passed. Indeed, we always (anti) symmetrize states only if it makes a difference for predictions. The symmetry postulate (like in classical mechanics) doesn't make a difference if states are orthogonal (which is the case is wave packets don't overlap). If we consider an isolated n-particle system, we consequently (anti) symmetrize only the state of this system and do not consider the rest of the universe. By doing this, we, in fact, started out by distinguishing the particles in the system from the particles in the rest of the universe! I think, most advocates of the h-particle concept don't have a problem with this---but they should have! Although it does not make a difference for the empirical results of QM, it does make a difference for the conceptual status of the h-particles, whether or not you choose to (anti) symmetrize.

Of course, q-particles are not like classical particles. We cannot define their states uniquely.
This is a peculiarity which is connected to 'wave-particle duality', not to the symmetry postulate!

**Acknowledgement**
I would like to thank Dennis Dieks for his valuable annotations to previous versions of this text. They have made a major difference.